\documentclass[prd,superscriptaddress,unsortedaddress,twocolumn,showpacs,preprintnumbers,amsmath,amssymb]{revtex4}
\usepackage[colorlinks=true, pdfstartview=FitV, linkcolor=red, citecolor=blue, urlcolor=blue]{hyperref}
\usepackage{graphicx}
\usepackage{bm}
\usepackage{ulem}

\graphicspath{{./fig/}}

\newcommand\Z{{\cal Z}}

\renewcommand\sout{\bgroup \color{red} \ULdepth=-.5ex \ULset}

\begin{document}
\preprint{KUNS-2681, YITP-17-54}

\title{Toward solving the sign problem with path optimization method}

\author{Yuto Mori}
\email[]{mori.yuto.47z@st.kyoto-u.ac.jp}
\affiliation{Department of Physics, Faculty of Science,
Kyoto University, Kyoto 606-8502, Japan}

\author{Kouji Kashiwa}
\email[]{kouji.kashiwa@yukawa.kyoto-u.ac.jp}
\affiliation{Yukawa Institute for Theoretical Physics,
Kyoto University, Kyoto 606-8502, Japan}

\author{Akira Ohnishi}
\email[]{ohnishi@yukawa.kyoto-u.ac.jp}
\affiliation{Yukawa Institute for Theoretical Physics,
Kyoto University, Kyoto 606-8502, Japan}

\begin{abstract}
We propose a new approach to circumvent the sign problem
 in which the integration path is optimized to control the sign problem.
We give a trial function specifying the integration path in
 the complex plane and tune it
 to optimize the cost function which represents the seriousness
 of the sign problem.
We call it the path optimization method.
In this method, we do not need to solve the gradient flow required
 in the Lefschetz-thimble method
 and then the construction of the integration-path contour
arrives at the optimization problem where several efficient methods can
be applied.
In a simple model with a serious sign problem, 
 the path optimization method is demonstrated to work
 well; the residual sign problem is resolved and precise results
 can be
 obtained even in the region where the global sign problem is serious.
\end{abstract}

\maketitle

{\it Introduction ---}~~~
The sign problem induced by the oscillating
Boltzmann weight of the partition function in the numerical integration
for various quantum systems is serious obstruction in the computational
science; see
Ref.~\cite{deForcrand:2010ys} for a review.
Particularly, the sign problem attracts much more attention recently in
QCD because some new approaches to circumvent the sign problem have
been proposed and applied.

Recent promising approaches to evade the sign problem include the complex
Langevin method~\cite{Parisi:1980ys,Parisi:1984cs,Aarts:2009uq}
and the Lefschetz-thimble path-integral
method~\cite{Witten:2010cx,Cristoforetti:2012su,Fujii:2013srak}.
The complex Langevin method is based on the stochastic quantization and
then we are free from the complex weight.
Therefore, the sign problem does not appear, but it is well known that
the complex Langevin method sometimes provides us
wrong results when the drift term shows singular behavior in the
Langevin-time evolution~\cite{Nishimura:2015pba}.
In comparison,
the Lefschetz-thimble path-integral method is based on the
Picard-Lefschetz theory~\cite{pham1983vanishing} and thus it is within
the standard path-integral formulation.
In this method, we construct the new integration-path contour which is
so called the Lefschetz thimbles by solving
the gradient flow starting from fixed points.
Then, the partition function can be decomposed into the summation over
contributions of relevant Lefschetz thimbles which can be determined
from the crossing behavior of conjugate gradient flows with the
original integration-path contour.
On each Lefschetz thimble, the imaginary part of the action is constant
and thus the sign problem seems to be resolved, but there are two
remnants of the original sign problem.
First one is the global sign problem:
In the summation process of relevant Lefschetz thimbles,
the cancellation can appear because imaginary parts of the action
are constant on one Lefschetz thimble
but have different values on different thimbles.
The other is the residual sign problem; it comes from the Jacobian
generated by the bending structure of the new integration path.
Recently, one more serious problem in the Lefschetz-thimble path-integral
method has been discussed which is so called
the singularity problem: There are singular points and cuts on the
complexified variables of integration if the action has the square root
and/or the logarithm, explicitly and implicitly~\cite{Mori:2017zyl}.
These singularities obstruct to draw continuous Lefschetz-thimbles in
the numerical calculation of gradient flows.

In this article, we propose a new method which we call the {\it
path optimization method} to attack the sign
and singularity problem.
This method is motivated by
the Lefschetz-thimble path-integral method.
The main idea is the modification of the path-integral contour by
minimizing the suitable {\it cost function}
which reflects the seriousness of the sign problem.
This means that the evading the sign problem arrives at the
optimization problem.
This fact becomes the strong advantage of this
method because the optimization problem is well explored in the
computational science and thus we may use several efficient methods such
as the machine learning in the optimization process~\cite{Mori:2017nwj}.
The path optimization method is demonstrated
in the simple model with the serious sign problem where the
complex Langevin method can fail.\\

{\it Cost and trial functions ---}~~~
In the path optimization method,
the function which is so called the {\it cost function} plays a
crucial role to construct the new integration-path contour
on which the sign problem is controllable.
The cost function is related with
the seriousness of the sign problem with
weakened weight cancellation by minimizing the function.
In this article, we use the following cost function;
\begin{align}
{\cal F}[z(t)]
&= \frac{1}{2} \int dt
   |e^{i \theta(t)} - e^{i \theta _0}|^2 \times |J(t) e^{-S(z(t))}|,
\label{Eq:cost_function}
\end{align}
with
\begin{align}
 e^{i \theta(t)} &= \frac{J(t)e^{-S(z(t))} }{|J(t)e^{-S(z(t))}|}, ~~~~
 e^{i \theta_0}   = \frac{\Z}{|\Z|},
\end{align}
where $z$ is the complexified variables of integration, $\Z$ is
the partition function and $J(t) = dz/dt$.
This function can be expressed by using the average
phase factor as
\begin{align}
\frac{{\cal F}}{|\Z|}
= |\langle e^{i\theta} \rangle _{pq}|^{-1} - 1,
\end{align}
where
\begin{align}
\langle {\cal O} \rangle _{pq} &\equiv \frac{\int dt {\cal O}|Je^{-S}|}{\int dt |Je^{-S}|}.
\label{Eq:evpq}
\end{align}
It should be noted that the choice of the cost function is
not unique and thus we can freely change or extend it as long as the
function reflects the seriousness of the sign problem.

To perform the optimization of Eq.~(\ref{Eq:cost_function}), we
need the trial function to specify the integration-path contour.
One simple way to prepare the trial function is using the complete set,
${\cal H}_m$, as
\begin{align}
z(t) = x(t) + iy(t),~~~~
 \begin{cases}
   x(t) &= \sum_m c_{x,m} {\cal H}_m (t) + t, \\
   y(t) &= \sum_m c_{y,m} {\cal H}_m (t),
 \end{cases}
\label{Eq:op}
\end{align}
with imposing the conditions,
$x(\pm \infty)=\pm \infty,~|y(\pm \infty)|<\infty$.
If the integrand, $\exp(-S)$, is analytic and suppressed rapidly enough with $|x| \to \infty$,
the integrals on the original and modified paths are the same owing to
Cauchy's integral theorem as long as the path does not go across singular points of $\exp(-S)$.
It should be noted that we do not need to care the singular point of $S$
if it is not a singular point of $\exp(-S)$.
We can extend this trial function to more complicated form
by performing the feature engineering or the machine learning~\cite{Mori:2017nwj}.

{\it Example ---}~~~
In this article, we consider the following partition
function as an example to demonstrate the path optimization method.
The actual form of partition function~\cite{Nishimura:2015pba} is
\begin{align}
\Z_p &= \int dx ~(x + i\alpha)^p e^{-\frac{x^2}{2}},
\end{align}
where $\alpha$ and $p$ are input parameters and $p$ is a positive integer.
The analytic result of $\Z_p$ can be obtained from the recurrence formula;
\begin{align}
\Z_p &= i\alpha \Z_{p-1} + (p-1)\Z_{p-2},
\end{align}
and the expectation value of $x^2$ is expressed as
\begin{align}
 \langle x^2 \rangle _p &= \frac{\Z_{p+2} - 2i\alpha\Z_{p+1} - \alpha^2\Z_p}{\Z_p}.
\label{Eq:x2}
\end{align}

In the path optimization method, we need to care the singular points of $\exp(-S)$.
In the present action, the relevant singular points exist
at ${\rm Im}~z \to \pm \infty$.
It should be noted that the singular point in $S$ at $z=-i\alpha$
does not cause any trouble.
The factor of integrand $(z+i\alpha)^p$
leads to the action term of $-p\log(z+i\alpha)$
and causes the singular drift term,
the drift term which diverges at $z=-i\alpha$,
in the complex Langevin method~\cite{Nishimura:2015pba}.
However,  $\exp(-S)$ is analytic at this point
as long as $p$ is taken to be a positive integer,
then it is not necessary to care in the path optimization method.
Nevertheless, "singular point" indicates this zero point
in the following discussions.

In the actual optimization, we use a simplified version of
Eq.~(\ref{Eq:op}) based on the Gaussian function;
\begin{align}
x(t) &= t, \\
y(t) &= c_1 \exp \Bigl(-\frac{c_2^2 t^2}{2} \Bigr) + c_3.
\end{align}
The optimization is numerically performed using the steepest descent
method, $dc_i/d\tau = -\partial {\cal F}/\partial c_i$,
and the integration is performed using the double exponential formula.

\begin{figure}[htb]
\begin{center}
 \includegraphics[width=0.4\textwidth]{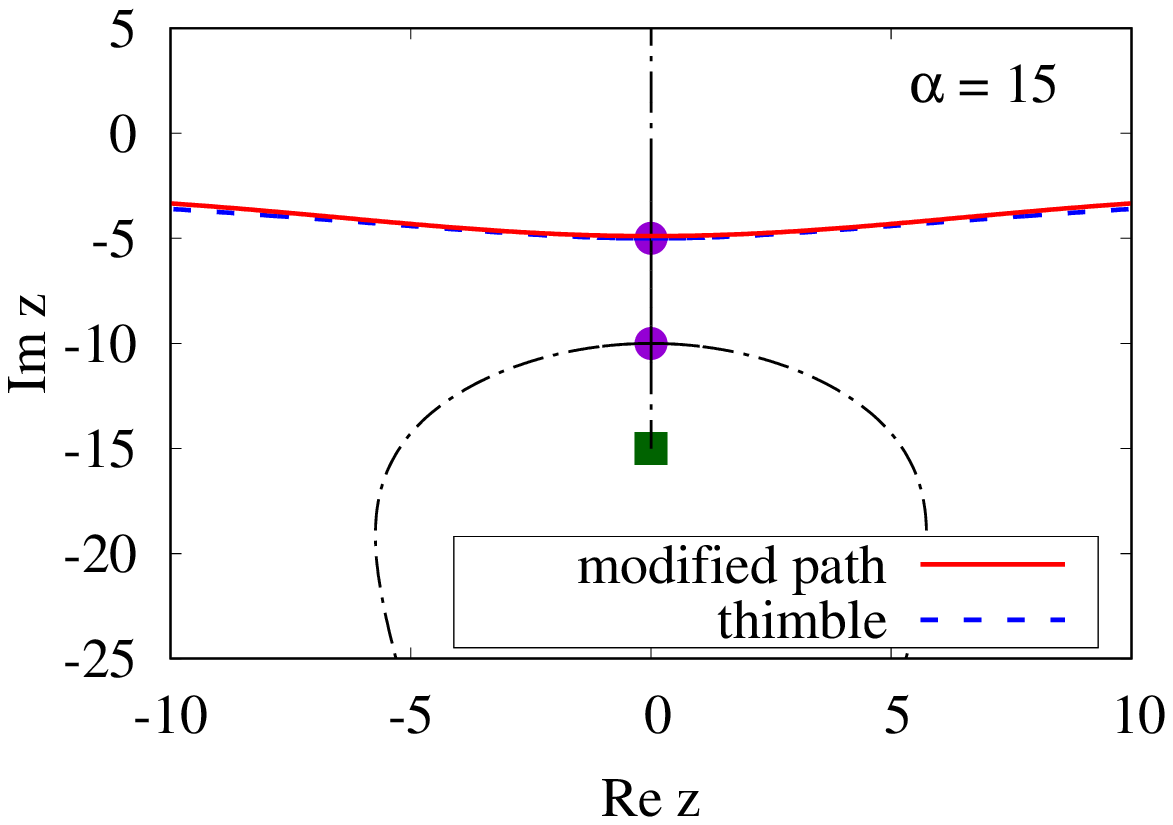}
\end{center}
\begin{center}
 \includegraphics[width=0.4\textwidth]{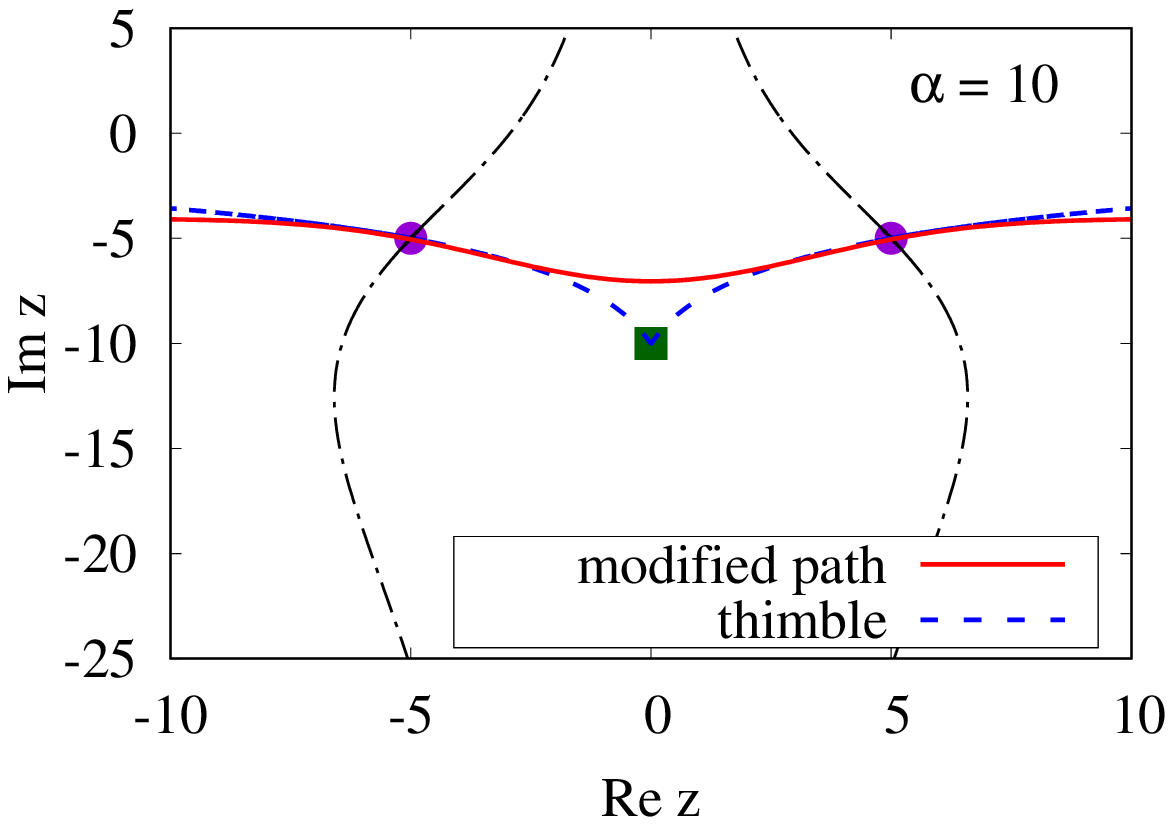}
\end{center}
 \caption{
 Modified integration path by optimizing
 Eq.~(\ref{Eq:cost_function})
 and the Lefschetz thimbles for $p=50, \alpha=15,10$.
 Closed circle (square) point shows the fixed (singular) point.
 Dot-dashed lines are steepest ascent paths.
 }
\label{Fig:path}
\end{figure}

The optimized integration path in comparison with the Lefschetz thimble is
shown in Fig.~\ref{Fig:path}.
It can be seen that the two contours overlap in the vicinity of the
fixed point.
However, there are qualitative differences on
the thimble structure with varying $\alpha$.
In the case with $\alpha=10$, Lefschetz thimbles
terminate at the singular point unlike the case with
$\alpha=15$ and then the optimized integration path
approaches the singular point.

\begin{figure}[htb]
\begin{center}
 \includegraphics[width=0.4\textwidth]{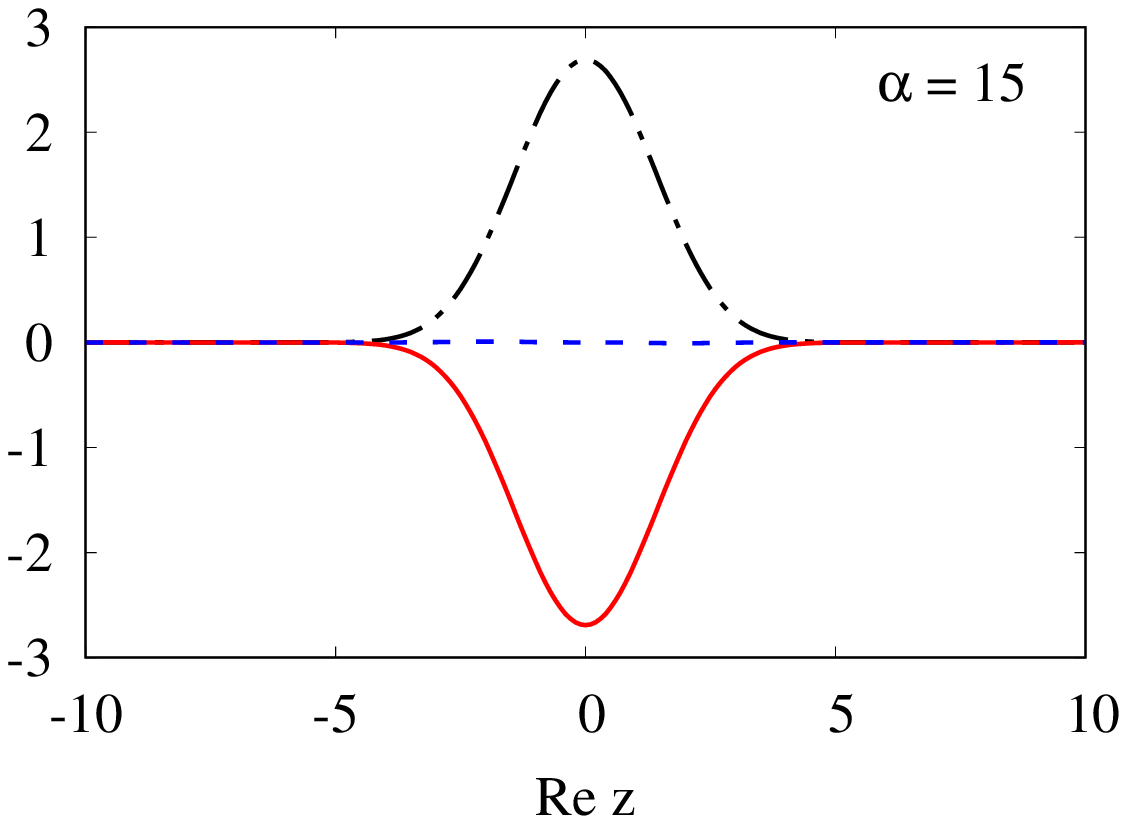}
\end{center}
\begin{center}
 \includegraphics[width=0.4\textwidth]{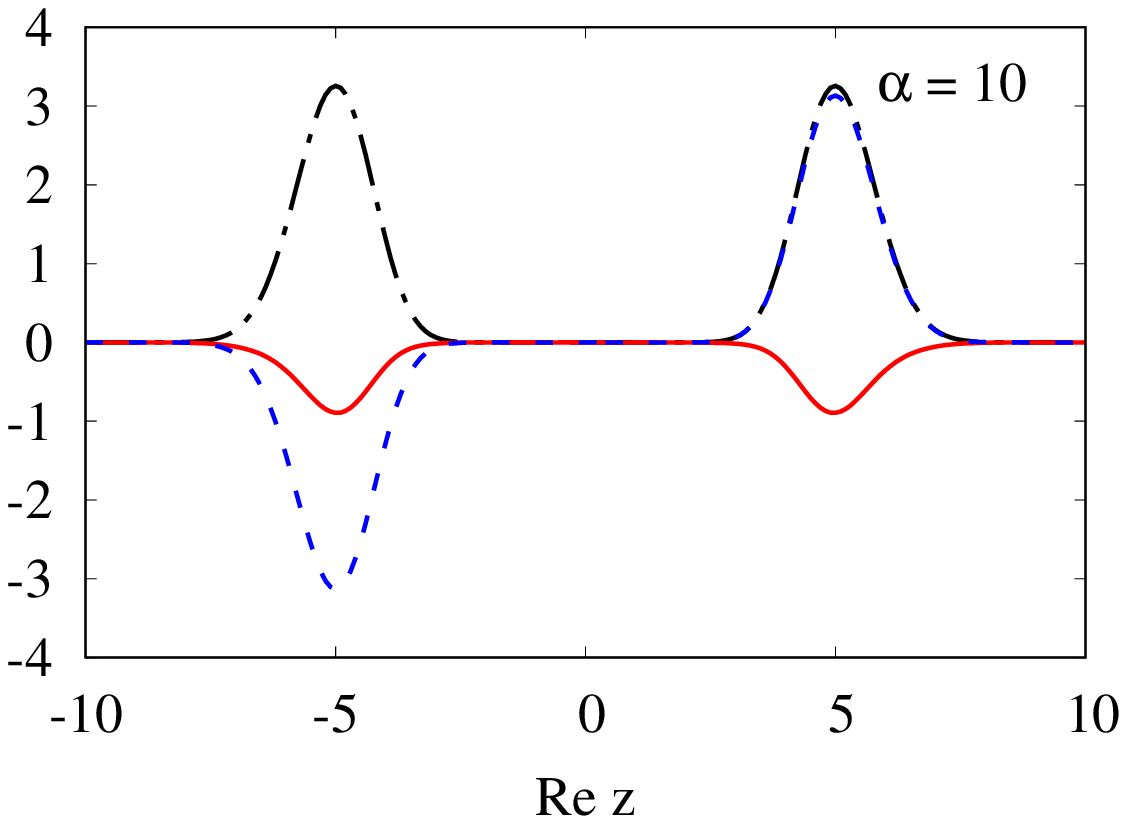}
\end{center}
 \caption{
 Boltzmann weight on the modified integration path with $p=50$,
 $\alpha=15,10$.
 Solid (dashed) line represents real (imaginary) part of $Je^{-S}$.
 The dot-dashed line indicates $|Je^{-S}|$.
 Where, the amplitudes is normalized by factor $10^{55}$ with $\alpha=15$, and $10^{42}$ with $\alpha=10$.
 }
\label{Fig:boltzmann}
\end{figure}

Figure~\ref{Fig:boltzmann} shows $Je^{-S}$ on the
optimized integration path.
We can see that there is the large probability
distribution ($W(t) \sim|Je^{-S}|$) with almost the constant phase
near the fixed point.
Therefore, Monte-Carlo sampling works with $\alpha = 15$.
In comparison, $W(t)$ has two peaks in the case with
$\alpha=10$.
The sign of $\mathrm{Im}~Je^{-S}$ are opposite at both peaks
and thus there are serious cancellations between them when we take
into account both peaks to the integration.
If we take into account only one peak,
the wrong result comes up.
The cancellations are induced 
by the singular point when the
optimized integration path approaches to it: The Boltzmann weight becomes zero
at the singular point and thus the sign of the Boltzmann weight can be
easily flipped near the singular point.
In the present case, there is exact parity symmetry
between ${\rm Re}~z$ and $-{\rm Re}~z$ and thus
the cancellation is very serious.
This cancellation reflects the hidden Lefschetz thimble
structure behind the path optimization method.

\begin{figure}[htb]
\begin{center}
 \includegraphics[width=0.4\textwidth]{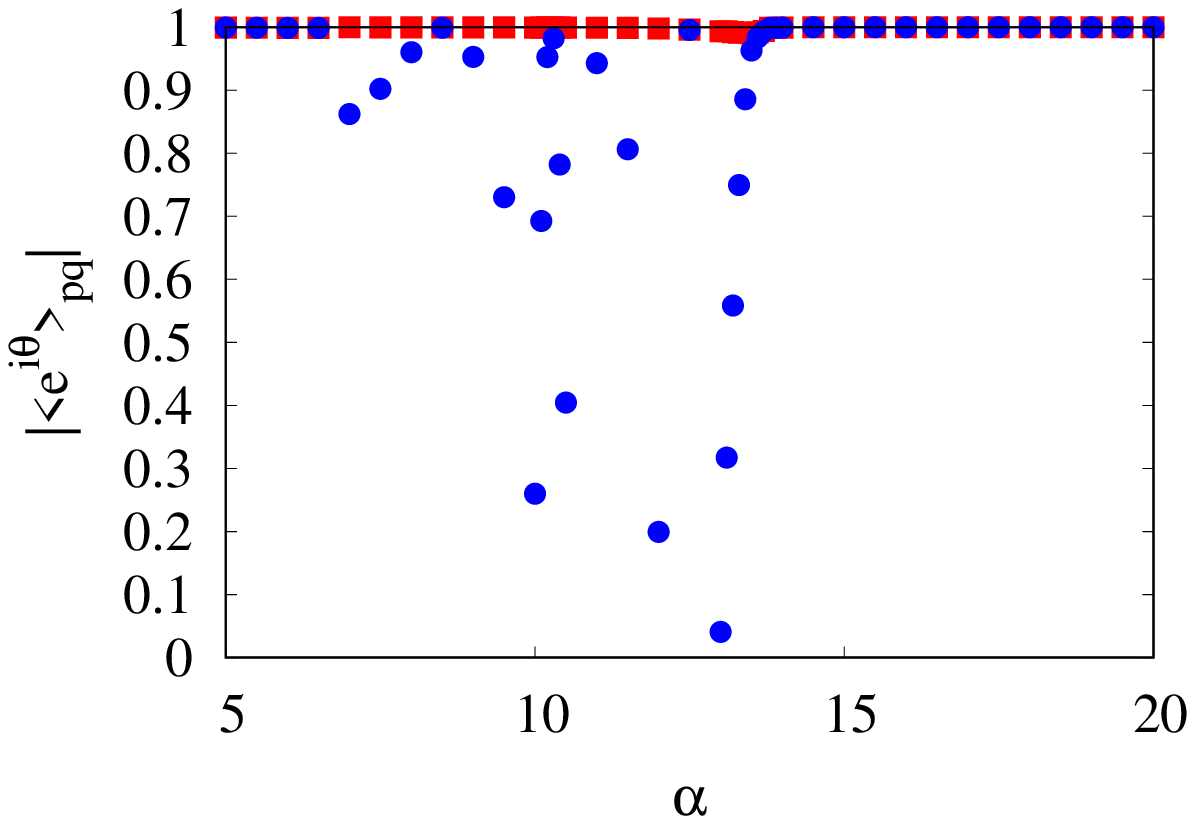}
\end{center}
 \caption{
 The average phase 
 factor with $p=50$.
 Closed circle shows the expectation in $\mathrm{Re}~z \in (-\infty,\infty)$, 
  and square point in $\mathrm{Re}~z \in [0,\infty)$.
 The expectation value is calculated by numerical integration.
 }
\label{Fig:error}
\end{figure}

The optimized average phase factor is shown in Fig.~\ref{Fig:error}.
From the difference between the full calculation and the calculation in
the $\mathrm{Re}~z \in [0,\infty)$ range, we can clarify the seriousness
of the global sign problem.
In the case with $\alpha \gtrsim 14$, we can see that the sign problem
can be solved because the path is represented by one thimble.
In the case with $\alpha \lesssim 14$, contributions from
the two thimbles cancel with each other.
In the path optimization method, we can resolve the residual sign
problem, but not the global sign problem.
This problem also exists in the ordinary and generalized Lefschetz
thimble methods~\cite{Alexandru:2015xva}.

\begin{figure}[htb]
\begin{center}
 \includegraphics[width=0.4\textwidth]{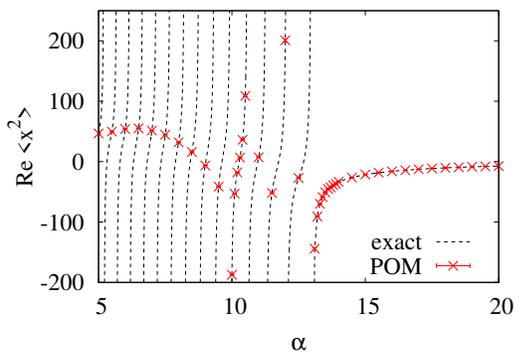} 
\end{center}
 \caption{
 The expectation value of $x^2$ with $p=50$. Errors are estimated by
 Jackknife method.
 }
\label{Fig:expectation}
\end{figure}

Figure~\ref{Fig:expectation} shows the expectation
value of $x^2$ in the hybrid Monte-Carlo method
on the modified integration path.
We calculate the expectation value in Eq.~(\ref{Eq:evpq}) by using
the reflection symmetry of the Boltzmann weight, $W(t)=W(-t)$, in this
setting.
This treatment replicates the parallel tempering algorithm which has
been applied to the generalized Lefschetz-thimble path-integral
method~\cite{Fukuma:2017fjq,Alexandru:2017oyw}.
The results well agree with the analytic results (\ref{Eq:x2}).
Readers can find the calculation of this model by using the complex
Langevin method in Ref.~\cite{Nishimura:2015pba}.

Because the path optimization method is a variational method,
we can restrict the variational space to represent
integration path.
This will be useful for multiple integral problems such as quantum field theory.
In addition, the path optimization method has large extensibility because
we can freely replace or extend the cost function as long as it
reflects the seriousness of the sign problem.
These points are important advantages of the path optimization method.
We leave the actual extension of the trial function and
the cost function as a future work
since this study is the first attempt to demonstrate our method.

{\it Summary ---}~~~
In this article, we have proposed a new approach to
circumvent the sign problem which is motivated by
the Lefschetz-thimble path-integral method.
We call it the {\it path optimization method}.
In the method, the new integration path is constructed in the
plane of complexified variables of integration by minimizing the {\it
cost function}.
The cost function is set to reflect the
seriousness of the sign problem.
The actual optimization of the
integration path is carried out by
using the trial function.

We have demonstrated the path optimization method by using the simple
model with the serious sign problem proposed in
Ref.~\cite{Nishimura:2015pba}.
In the path optimization method, we can resolve the residual sign
problem which appears in
the ordinary and generalized Lefschetz-thimble path-integral methods.
But, at least on our present choice of the cost function and
in the present setting,
the global sign problem cannot be resolved.
However, we can well reproduce the exact results by using the path
optimization method in the wide range of the model parameter
space.

Finally, we summarize advantages of the path optimization method:
\begin{enumerate}
  \item No residual sign problem.
  \item Applicability of various efficient methods to the
	optimization process.
  \item Controllability of the singularity problem.
  \item Large extensibility of the cost function.
\end{enumerate}
Possible disadvantage may be the numerical cost. In the complex system,
the sign-problem weakened integration path
is expected to have a very complicated shape.
Therefore, we should check which optimization method is suitable or not
step by step in the future.

\vspace{2mm}
{\it Acknowledgments:}
A.O. is supported in part by the Grants-in-Aid for Scientific Research
 from JSPS (Nos. 15K05079, 15H03663, 16K05350),
the Grants-in-Aid for Scientific
 Research on Innovative Areas from MEXT (Nos. 24105001, 24105008),
 and by the Yukawa International Program for Quark-hadron
 Sciences (YIPQS).

\bibliography{ref.bib}

\begin{thebibliography}{14}
\expandafter\ifx\csname natexlab\endcsname\relax\def\natexlab#1{#1}\fi
\expandafter\ifx\csname bibnamefont\endcsname\relax
  \def\bibnamefont#1{#1}\fi
\expandafter\ifx\csname bibfnamefont\endcsname\relax
  \def\bibfnamefont#1{#1}\fi
\expandafter\ifx\csname citenamefont\endcsname\relax
  \def\citenamefont#1{#1}\fi
\expandafter\ifx\csname url\endcsname\relax
  \def\url#1{\texttt{#1}}\fi
\expandafter\ifx\csname urlprefix\endcsname\relax\def\urlprefix{URL }\fi
\providecommand{\bibinfo}[2]{#2}
\providecommand{\eprint}[2][]{\url{#2}}

\bibitem[{\citenamefont{de~Forcrand}(2009)}]{deForcrand:2010ys}
\bibinfo{author}{\bibfnamefont{P.}~\bibnamefont{de~Forcrand}},
  \bibinfo{journal}{PoS} \textbf{\bibinfo{volume}{LAT2009}},
  \bibinfo{pages}{010} (\bibinfo{year}{2009}), \eprint{1005.0539}.

\bibitem[{\citenamefont{Parisi and Wu}(1981)}]{Parisi:1980ys}
\bibinfo{author}{\bibfnamefont{G.}~\bibnamefont{Parisi}} \bibnamefont{and}
  \bibinfo{author}{\bibfnamefont{Y.-s.} \bibnamefont{Wu}},
  \bibinfo{journal}{Sci.Sin.} \textbf{\bibinfo{volume}{24}},
  \bibinfo{pages}{483} (\bibinfo{year}{1981}).

\bibitem[{\citenamefont{Parisi}(1983)}]{Parisi:1984cs}
\bibinfo{author}{\bibfnamefont{G.}~\bibnamefont{Parisi}},
  \bibinfo{journal}{Phys.Lett.} \textbf{\bibinfo{volume}{B131}},
  \bibinfo{pages}{393} (\bibinfo{year}{1983}).

\bibitem[{\citenamefont{Aarts et~al.}(2010)\citenamefont{Aarts, Seiler, and
  Stamatescu}}]{Aarts:2009uq}
\bibinfo{author}{\bibfnamefont{G.}~\bibnamefont{Aarts}},
  \bibinfo{author}{\bibfnamefont{E.}~\bibnamefont{Seiler}}, \bibnamefont{and}
  \bibinfo{author}{\bibfnamefont{I.-O.} \bibnamefont{Stamatescu}},
  \bibinfo{journal}{Phys. Rev.} \textbf{\bibinfo{volume}{D81}},
  \bibinfo{pages}{054508} (\bibinfo{year}{2010}), \eprint{0912.3360}.

\bibitem[{\citenamefont{Witten}(2011)}]{Witten:2010cx}
\bibinfo{author}{\bibfnamefont{E.}~\bibnamefont{Witten}},
  \bibinfo{journal}{AMS/IP Stud. Adv. Math.} \textbf{\bibinfo{volume}{50}},
  \bibinfo{pages}{347} (\bibinfo{year}{2011}), \eprint{1001.2933}.

\bibitem[{\citenamefont{Cristoforetti et~al.}(2012)\citenamefont{Cristoforetti,
  Di~Renzo, and Scorzato}}]{Cristoforetti:2012su}
\bibinfo{author}{\bibfnamefont{M.}~\bibnamefont{Cristoforetti}},
  \bibinfo{author}{\bibfnamefont{F.}~\bibnamefont{Di~Renzo}}, \bibnamefont{and}
  \bibinfo{author}{\bibfnamefont{L.}~\bibnamefont{Scorzato}}
  (\bibinfo{collaboration}{AuroraScience Collaboration}),
  \bibinfo{journal}{Phys.Rev.} \textbf{\bibinfo{volume}{D86}},
  \bibinfo{pages}{074506} (\bibinfo{year}{2012}), \eprint{1205.3996}.

\bibitem[{\citenamefont{Fujii et~al.}(2013)\citenamefont{Fujii, Honda, Kato,
  Kikukawa, Komatsu et~al.}}]{Fujii:2013srak}
\bibinfo{author}{\bibfnamefont{H.}~\bibnamefont{Fujii}},
  \bibinfo{author}{\bibfnamefont{D.}~\bibnamefont{Honda}},
  \bibinfo{author}{\bibfnamefont{M.}~\bibnamefont{Kato}},
  \bibinfo{author}{\bibfnamefont{Y.}~\bibnamefont{Kikukawa}},
  \bibinfo{author}{\bibfnamefont{S.}~\bibnamefont{Komatsu}},
  \bibnamefont{et~al.}, \bibinfo{journal}{JHEP}
  \textbf{\bibinfo{volume}{1310}}, \bibinfo{pages}{147} (\bibinfo{year}{2013}),
  \eprint{1309.4371}.

\bibitem[{\citenamefont{Nishimura and Shimasaki}(2015)}]{Nishimura:2015pba}
\bibinfo{author}{\bibfnamefont{J.}~\bibnamefont{Nishimura}} \bibnamefont{and}
  \bibinfo{author}{\bibfnamefont{S.}~\bibnamefont{Shimasaki}},
  \bibinfo{journal}{Phys. Rev.} \textbf{\bibinfo{volume}{D92}},
  \bibinfo{pages}{011501} (\bibinfo{year}{2015}), \eprint{1504.08359}.

\bibitem[{\citenamefont{Pham}(1983)}]{pham1983vanishing}
\bibinfo{author}{\bibfnamefont{F.}~\bibnamefont{Pham}}, in
  \emph{\bibinfo{booktitle}{Proc. Symp. Pure Math}} (\bibinfo{publisher}{AMS},
  \bibinfo{year}{1983}), \bibinfo{number}{40}, pp. \bibinfo{pages}{319--333}.

\bibitem[{\citenamefont{Mori et~al.}(2017{\natexlab{a}})\citenamefont{Mori,
  Kashiwa, and Ohnishi}}]{Mori:2017zyl}
\bibinfo{author}{\bibfnamefont{Y.}~\bibnamefont{Mori}},
  \bibinfo{author}{\bibfnamefont{K.}~\bibnamefont{Kashiwa}}, \bibnamefont{and}
  \bibinfo{author}{\bibfnamefont{A.}~\bibnamefont{Ohnishi}}
  (\bibinfo{year}{2017}{\natexlab{a}}), \eprint{1705.03646}.

\bibitem[{\citenamefont{Mori et~al.}(2017{\natexlab{b}})\citenamefont{Mori,
  Kashiwa, and Ohnishi}}]{Mori:2017nwj}
\bibinfo{author}{\bibfnamefont{Y.}~\bibnamefont{Mori}},
  \bibinfo{author}{\bibfnamefont{K.}~\bibnamefont{Kashiwa}}, \bibnamefont{and}
  \bibinfo{author}{\bibfnamefont{A.}~\bibnamefont{Ohnishi}}
  (\bibinfo{year}{2017}{\natexlab{b}}), \eprint{1709.03208}.

\bibitem[{\citenamefont{Alexandru et~al.}(2016)\citenamefont{Alexandru, Basar,
  and Bedaque}}]{Alexandru:2015xva}
\bibinfo{author}{\bibfnamefont{A.}~\bibnamefont{Alexandru}},
  \bibinfo{author}{\bibfnamefont{G.}~\bibnamefont{Basar}}, \bibnamefont{and}
  \bibinfo{author}{\bibfnamefont{P.}~\bibnamefont{Bedaque}},
  \bibinfo{journal}{Phys. Rev.} \textbf{\bibinfo{volume}{D93}},
  \bibinfo{pages}{014504} (\bibinfo{year}{2016}), \eprint{1510.03258}.

\bibitem[{\citenamefont{Fukuma and Umeda}(2017)}]{Fukuma:2017fjq}
\bibinfo{author}{\bibfnamefont{M.}~\bibnamefont{Fukuma}} \bibnamefont{and}
  \bibinfo{author}{\bibfnamefont{N.}~\bibnamefont{Umeda}}
  (\bibinfo{year}{2017}), \eprint{1703.00861}.

\bibitem[{\citenamefont{Alexandru et~al.}(2017)\citenamefont{Alexandru, Basar,
  Bedaque, and Warrington}}]{Alexandru:2017oyw}
\bibinfo{author}{\bibfnamefont{A.}~\bibnamefont{Alexandru}},
  \bibinfo{author}{\bibfnamefont{G.}~\bibnamefont{Basar}},
  \bibinfo{author}{\bibfnamefont{P.~F.} \bibnamefont{Bedaque}},
  \bibnamefont{and} \bibinfo{author}{\bibfnamefont{N.~C.}
  \bibnamefont{Warrington}} (\bibinfo{year}{2017}), \eprint{1703.02414}.

\end{thebibliography}

\end{document}